# Extended Quantum Anomalous Hall States in Graphene/hBN Moiré Superlattices

Zhengguang Lu[1]†, Tonghang Han[1]†, Yuxuan Yao[1]†, Jixiang Yang[1], Junseok Seo[1], Lihan Shi[1], Shenyong Ye[1], Kenji Watanabe[2], Takashi Taniguchi[3], Long Ju[1]*

[1]Department of Physics, Massachusetts Institute of Technology, Cambridge, MA, USA.

[2]Research Center for Electronic and Optical Materials, National Institute for Materials Science, 1-1 Namiki, Tsukuba 305-0044, Japan

[3]Research Center for Materials Nanoarchitectonics, National Institute for Materials Science, 1-1 Namiki, Tsukuba 305-0044, Japan

*Corresponding author. Email: longju@mit.edu †These authors contributed equally to this work.

**Electrons in topological flat bands can form novel topological states driven by the correlation effects. The penta-layer rhombohedral graphene/hBN moiré superlattice has been shown to host fractional quantum anomalous Hall effect (FQAHE) at ~400 mK[1], triggering discussions around the underlying mechanism and the role of moiré effects[2–6]. In particular, novel electron crystal states with non-trivial topology have been proposed[3,4,7–14]. Here we report DC electrical transport measurement in rhombohedral penta- and tetra-layer graphene/hBN moiré superlattices at electronic temperatures down to ~40 mK. We observed two more FQAH states in the penta-layer devices than previously reported. In a new tetra-layer device, we observed FQAHE at filling factors $v$ = 3/5 and 2/3 at 300 mK. With a small bias current and the lowest temperature, we observed a new extended quantum anomalous Hall (EQAH) state and magnetic hysteresis, where $R_{xy}$ = h/e² and vanishing $R_{xx}$ span a wide range of moiré filling factor $v$ from 0.5 to up to 1.3. By increasing the temperature or current, FQAHE can be recovered---suggesting the break-down of the EQAH states and a phase transition into the fractional quantum Hall liquid[15–17]. Furthermore, we observed displacement field-induced quantum phase transitions from the EQAH states to Fermi liquid, FQAH liquid and the likely composite Fermi liquid. Our observation establishes a new topological phase of electrons with quantized Hall resistance at zero magnetic field, and enriches the emergent quantum phenomena in materials with topological flat bands.**

Flat electronic bands with quantum geometry have been a rich platform to explore emergent quantum phenomena driven by intertwined correlation and topology effects. In the heterostructures of two-dimensional materials, such bands can be engineered with great flexibilities in structures and tuned *in situ* through dual-gating. Consequently, a rich spectrum of quantum phases of matter has been discovered in these synthetic quantum materials. In the example of rhombohedral graphene-based heterostructures, gate-tunable topological flat bands stem from the highly ordered crystalline structure and additional proximity effects due to neighboring layers[18–36]. It has been demonstrated to host correlated insulators[19,20,23,25,26], orbital multiferroicity[24] and superconductivity[22,31]. In particular, the penta-layer rhombohedral graphene/hBN moiré superlattice has been shown to host the fractional quantum anomalous Hall effect (FQAHE) with six fractional states[1]. The underlying mechanism of integer and fractional QAHE in this system, however, is under debate[2–6]. This is partially due to the weak moiré potential experienced by electrons when exhibiting FQAHE, and the lack of an isolated moiré miniband in the single-particle picture. These apparent puzzles, the higher temperature scale of FQAHE than FQHE, and the highly tunable band structures in this ideal material platform leave a wide un-charted territory to explore other emergent quantum phenomena. Among a range of theoretical proposals to describe this system, some works suggest the possibility of a quantum anomalous Hall crystal (QAHC) that could exist under a weak moiré potential or even in the absence of moiré effects [3,4,10–14], which is yet to be experimentally tested or understood.

Here we report electrical transport properties of the rhombohedral graphene/hBN moiré superlattices at electronic temperatures down to ~40 mK. In Device 1&2, which are based on penta-layer graphene, we observed deeper dips in $R_{xx}$ at fractional fillings and two more FQAH states than previously reported[1]. In Device 3, which is based on tetra-layer graphene, we observed two fractional states that realized the FQAHE. These observations reinforced the FQAHE in this material system. At even lower temperatures and certain ranges of $D$, surprisingly, we observed quantized Hall resistance $R_{xy} = h/e^2$ in a wide range of moiré filling factor $v$ from 0.5 to up to 1.3 accompanied by vanishing longitudinal resistance $R_{xx}$. These states realize the integer quantum anomalous Hall effect[1,33,37–42] in an extended range of charge density and filling factors (thus named EQAH states), but are distinct from Chern insulators and re-entrant quantum Hall insulators in several aspects. At increased temperature or bias current, these states are broken down and replaced by certain FQAH states. We will discuss the nature of this newly emerged topological state in comparison to the QAHC, re-entrant quantum Hall insulators and Wigner crystals. The data presented in the main text is based on device 1&3, and additional data from device 2&3 are included in Extended Data Figures. Summary of the device information, including optical image, layer number and hBN alignment angle is shown in extended Figure 4.

**FQAHE in Penta- and Tetra-layer Graphene**

Figure 1a shows schematics of all three devices, in which FQAHE are observed when electrons are polarized to the layer furthest from the moiré superlattice. We have improved the filtering of the electronic noise and electron thermalization in our dilution refrigerator, and we believe that an electronic temperature of ~40 mK can be reached when the mixing chamber temperature is at the base of 10 mK.

We first examine the FQAHE in Device 1. Figure 1b shows $R_{xx}$ in Device 1, the penta-layer rhombohedral graphene/hBN moiré device, at $D/\varepsilon_0$ = 0.925 V/nm at 0.3, 0.2 and 0.1 K. Clear dips can be seen at fractional filling factors $v$ = 2/5, 3/7, 4/9, 4/7 and 3/5 as reported previously, while new dips can be identified at $v$ = 5/11 and 5/9 at 100 mK. At these filling factors, the value of $R_{xx}$ decreases by up to 2.5 times as the mixing chamber temperature is lowered from 0.3 K to 0.1 K.

The observations are aligned with the expectation of fractional quantum (anomalous) Hall states, in which the $R_{xx}$ decreases as the temperature is lowered and the thermal activation across the FQ(A)H gaps is suppressed. In addition, a lower electronic temperature indeed helps revealing more fractional states that have larger denominators and smaller energy gaps than previously reported. Based on the temperature dependence of $R_{xx}$, we think the lowest electronic temperature in the previous measurement on these two samples was between 0.2 and 0.3 K.

We further examine a newly made tetra-layer graphene/hBN moiré superlattice (Device 3), the optical image of which is shown in Fig. 1c. We have optimized the device design and fabrication so that all the six electrodes have good electrical contact to the channel at large $D$, and standard Hall bar measurement is done. At a temperature of 0.3 K, quantized anomalous Hall resistance $R_{xy} = h/ve^2$ and dips in $R_{xx}$ are clearly seen at $v$ = 3/5, 2/3 and 1, as shown in Fig. 1d. The value of $R_{xx}$ at $v$ = 2/3 is around 0.1 $h/e^2$, even smaller than that in Device 1&2. The $R_{xx}$ and $R_{xy}$ maps, and the Landau fan data of Device 3 are included as Extended Data Figure 6-8.

The data in Fig. 1d demonstrates the FQAHE in the tetra-layer graphene/hBN moiré superlattice. Although the graphene layer number is different from the penta-layer devices, the general phase diagram is similar to that of the latter (see Ref.[1] and Extended Data Figure 6). This observation agrees with theoretical calculations[2–4,6], indicating that rhombohedral graphene/hBN moiré superlattice is a family of materials hosting FQAHE. We note that the twist angles are slightly different in Device 1&2 (~0.77 deg) and Device 3 (~0.55 deg), which results in the difference between the charge densities corresponding to $v$ = 1. This difference might explain the difference in the number of fractional states observed in tetra- and penta-layer devices (2 vs 8). In general, the effect of twist angle is to be examined experimentally by systematically fabricating and measuring more devices.

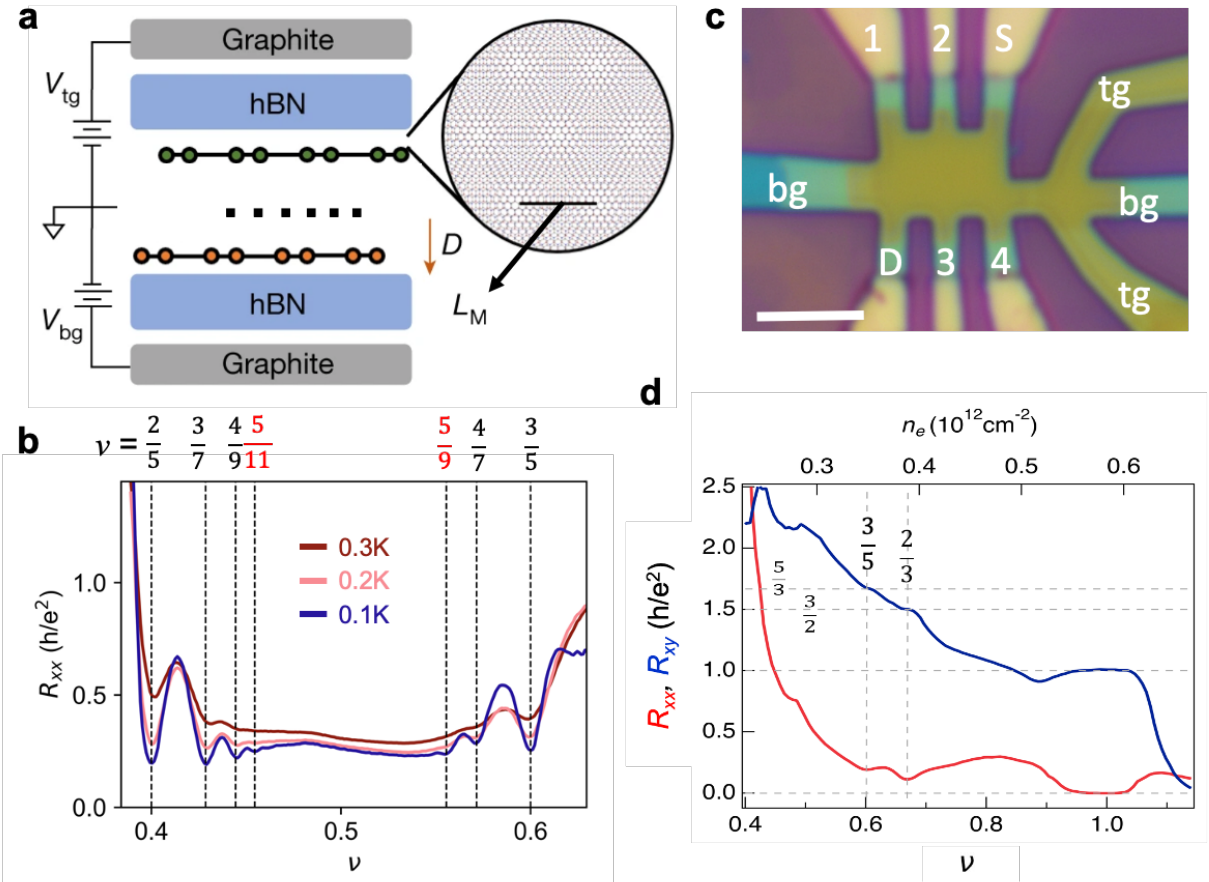

*Figure 1. Fractional quantum anomalous Hall effects in rhombohedral penta- and tetra-layer graphene/hBN moiré superlattice devices.* **a.** *Schematics of the device configuration and measurement conditions. In both penta- and tetra-layer devices, electrons are polarized to the layer that is furthest from the moiré superlattice when the FQAHE is observed.* **b.** *Temperature-dependent $R_{xx}$ of Device 1 the penta-layer device. Clear dips can be seen at $v$ = 2/5, 3/7, 4/9, 4/7 and 3/5 as reported previously, while new fractional states can be identified at $v$ = 5/11, and 5/9 at the lowest temperature. At a fixed $v$ corresponding to the FQAH state, $R_{xx}$ decreases as the temperature is lowered.* **c.** *Optical image of Device 3 the tetra-layer device. 'tg' and 'bg' correspond to top gate and bottom gate electrodes, respectively. Scale bar: 4 µm. We pass a current through 'S' & 'D', measure $V_{xx}$ between '1' & '2', and measure $V_{xy}$ between '2' & '3'.* **d.** *$R_{xx}$ and $R_{xy}$ in the tetra-layer device at 300 mK. Quantized $R_{xy}$ and dips in $R_{xx}$ can be seen at $v$ = 3/5 and 2/3---underscoring the universal FQAHE phenomena in rhombohedral graphene/hBN moiré superlattice systems. All data are extracted by symmetrizing and anti-symmetrizing data taken at out-of-plane magnetic fields $B = \pm 0.1$ T.*

## Extended Quantum Anomalous Hall States

Figure 2 shows the phase diagram measured in Device 1 in a large range of $v$ and gate-displacement field $D$. Three regions with uniform vanishing $R_{xx}$ and non-zero $R_{xy}$ values can be identified in Fig.

2a&b. The first one is a diamond-shaped region at $v = 1/2$[43], which is neighbored by a big second region covering from $v = 0.55$ to 0.9. The latter almost connects to the third region that covers from $v = 0.93$ to 1.03. These three regions compose a big anomalous Hall region, which sandwiches the FQAHE region together with the very insulating (and likely Wigner crystal[1]) region at low $v$ and $D$. Figure 2c&d reveal temperature-dependent quantitative details along the dashed lines in Fig. 2a&b, which cut through the three aforementioned regions. At the base temperature of 10 mK (mixing chamber temperature), $R_{xx}$ almost vanishes at all filling factors from $v = 1/2$ to 1. At the same time, $R_{xy}$ shows a wide plateau at $h/e^2$ covering the same range of $v$. There are two small gaps between the three regions, but the temperature dependence clearly suggests their destiny towards quantized values of $R_{xx} = 0$ and $R_{xy} = h/e^2$, should the electronic temperature be further lowered. When the temperature is increased, the values of $R_{xx}$ and $R_{xy}$ deviate from these quantized values and recovered to those reported previously at higher electronic temperatures[1].

In a tetra-layer rhombohedral graphene/hBN moiré superlattice device (see Extended Data Fig. 7), we observed similar states with quantized $R_{xy}$ and vanishing $R_{xx}$ in a wide range of $v$ similar to those shown in Fig. 2. There are two differences between the tetra-layer and penta-layer devices though: 1. In the former case, regions 2 and 3 with quantized $R_{xy} = h/e^2$ are merged into one region without a gap between them; 2. The quantized $R_{xy} = h/e^2$ state extends significantly beyond $v = 1$ and reaches $v = 1.3$.

The observation of quantized $R_{xy} = h/e^2$ and vanishing $R_{xx}$ across a wide range of filling factor ($v = 0.5$ to 1 for penta-layer graphene, $v = 0.5$ to 1.3 for tetra-layer graphene) clearly goes beyond the IQAHE and FQAHE physics picture[1,33,35,37–42]. We name these states as extended quantum anomalous Hall (EQAH) states. At $v = 3/5$ and $2/3$ and certain ranges of $D$, FQAH states are replaced by the EQAH states in both tetra-layer and penta-layer devices, suggesting the latter to be the true ground state at low enough temperatures for these ($v$, $D$) combinations in the second region. Even for the third region including $v = 1$, the plateau of $R_{xy}$ in Fig. 2d is twice wider at 10 mK than at 380 mK and includes more states at $v < 1$. At continuously varying $v$ from 0.5 to 1.3 (other than 3/5, 2/3 and 1), the quantization of $R_{xy}$ and $R_{xx}$ suggest that EQAH states bear a universal feature that is not necessarily related to the underlying moiré superlattice. Even for the smallest (region 1) among the three regions, the range of charge density and filling factor it covers is much broader than any FQAH state covers.

Figure 2e&f show the magnetic hysteresis behaviors of $R_{xx}$ and $R_{xy}$ at 10 mK at two representative positions on the phase diagram in the first and second quantized regions. Although with very different charge densities and filling factors, both states show the quantization of $R_{xy} = \pm h/e^2$ at saturation. Together with data in Fig. 2a-d, we conclude that the EQAH state is a new topological state that realizes the integer quantum anomalous Hall effect[37].

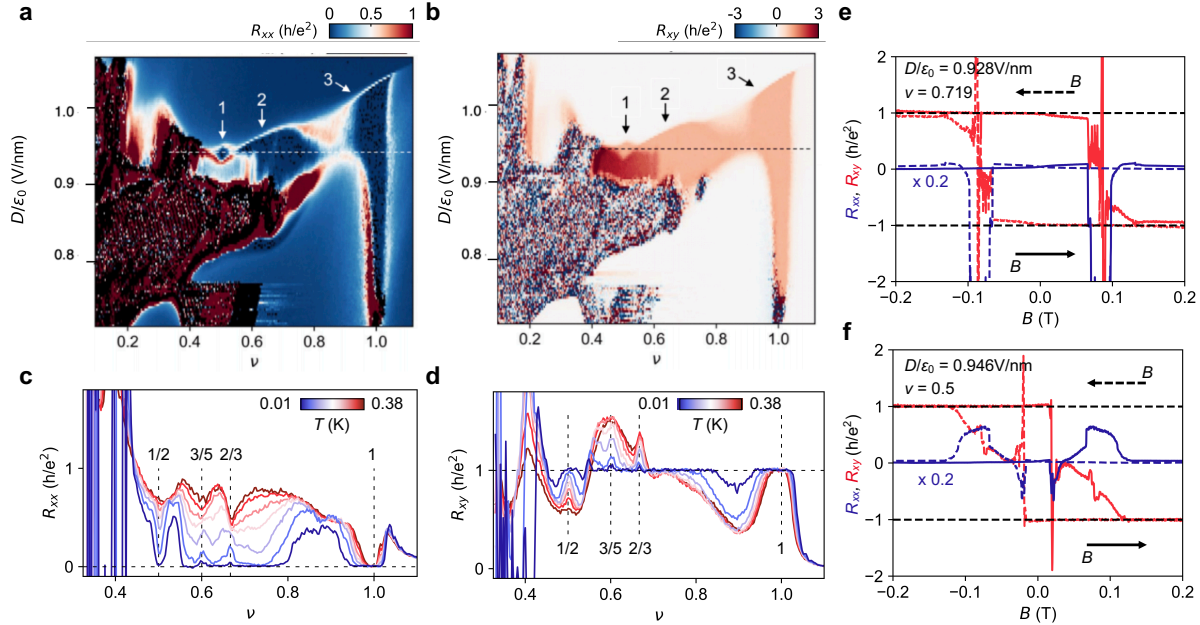

*Figure 2. Extended quantum anomalous Hall states in Device 1 the penta-layer device. **a&b.** Mapping of $R_{xx}$ and $R_{xy}$ in a large range of v and D at a mixing chamber temperature of 10 mK. Three regions (labeled by numbers and arrows) show quantized $R_{xy}$ at $h/e^2$ and vanishing $R_{xx}$, located at around v = 1/2, spanning between v = 0.55 and 0.83, as well as around v = 1. These three regions are almost connected into one big region with $R_{xy} = h/e^2$ that swamps the FQAH states at v > 1/2. **c&d.** Temperature-dependent $R_{xx}$ and $R_{xy}$ taken along the dashed line in a&b. From high temperature to low temperature, $R_{xx}$ decreases and the dips corresponding to FQAH states at v = 3/5 and 2/3 become small bumps, while the $R_{xy}$ approaches $h/e^2$ from values both below and above it. The plateau of $R_{xy}$ at $h/e^2$ spans almost from v = 1/2 to 1 at 10 mK, and the gap at v = 0.94 and 0.53 are on the trend of being closed as the temperature is lowered. **e&f.** Magnetic hysteresis scans of $R_{xx}$ and $R_{xy}$ at $(v,D/\varepsilon_0)$= (0.719, 0.928 V/nm) and (0.5, 0.946 V/nm) respectively at 10mK, featuring quantized $R_{xy} = h/e^2$ in both cases. All data are extracted by symmetrizing/anti-symmetrizing data taken at B = ± 0.1 T.*

## Current-Induced Break-down

Figure 3 shows the behaviors of EQAH states under varied current excitations. We apply a DC current through the sample and limit the AC current excitation to 50 pA to measure the corresponding differential resistances $R_{xy}$ and $R_{xx}$. Figure 3a&b show the results taken along the dashed lines in Fig. 2a&b at 10 mK. By increasing the DC current $I_{DC}$ from 0 to 2.3 nA, the wide plateaus of $R_{xy} = h/e^2$ and $R_{xx} = 0$ gradually disappear and eventually evolve into shapes that are generally similar to the curves at 380 mK in Fig. 2c&d and in our previous work[1].

Phenomenologically, increasing the DC current has the same effect of weakening the EQAH states as increasing the temperature does. At $I_{DC}$ = 2.3 nA, both $R_{xy}$ and $R_{xx}$ curves mimic in shape those in Fig. 2c&d at high temperatures, and the value of $R_{xy}$ recovers to that of the FQAHE at $v = 3/5$ and $2/3$. These observations suggest that the EQAH ground states is broken-down and replaced by the fractional quantum Hall liquids[15–17] under high DC current excitations.

However, there are also important differences between the $R_{xy}$ and $R_{xx}$ curves at elevated temperature and increased DC current. In Fig. 2c, $R_{xx}$ increases almost monotonically as the temperature is increased at all filling factors. In Fig. 3b, however, $R_{xx}$ first increases then decreases as the DC current increases. This non-monotonic trend is most obvious at $v = 0.6$-$0.8$ and can also be seen for other filling factors. Such non-monotonic dependence on the DC current can also be seen for $R_{xy}$ in Fig. 3a. We highlight such non-monotonic behaviors of $R_{xy}$ and $R_{xx}$ at two representative states that show $R_{xy} = h/e^2$ and $R_{xx} = 0$, as can be seen in Fig. 3c-f. The first state resides in the second region of quantized $R_{xy}$ while the second state resides inside the diamond-shaped first region of quantized $R_{xy}$. In both cases, $R_{xx}$ exhibits clear threshold behaviors: being zero at small current and showing a pair of peaks at a critical value of current. Correspondingly, $R_{xy}$ exhibits the quantized value of $h/e^2$ at small current and sudden changes at the same critical current as in the $R_{xx}$ curve. Such variations of $R_{xy}$ and $R_{xx}$ as a function of the DC current disappear at high temperatures, where both $R_{xy}$ and $R_{xx}$ remain constant.

Such threshold and non-monotonic behaviors of $R_{xy}$ and $R_{xx}$ indicate non-linear voltage-current relations. The observations are reminiscent of two other well-known examples of correlated electron ground states. Firstly, Fig. 3d resembles the differential resistance behavior in an s-wave superconductor: the zero-resistance Cooper pairs carry the current until hitting a critical value and a pair of resistance peaks emerge. In the case of EQAH states, the current is carried by the chiral edge states until reaching a critical value and the bulk transport starts to contribute. Secondly, such break-down behaviors suggested by the differential resistances are also reminiscent of the Chern insulators that exhibit integer quantum anomalous Hall effects[44–47]. In Chern insulators with the QAHE, the insulating bulk of the sample breaks down at high current/voltage, and transitions into a Fermi liquid state. Quantitatively, however, the breakdown of the Chern insulator and the EQAH states are quite different. In Extended Data Fig. 8, we compare the $R_{xx}$ vs $I_{DC}$ for the EQAH states and the Chern insulator at $v = 1$ in the same tetra-layer device. The threshold current in the EQAH state is 50 times smaller than that in the Chern insulator state of the same device---suggesting a different mechanism of break-down from previously reported for Chern insulators[44,45].

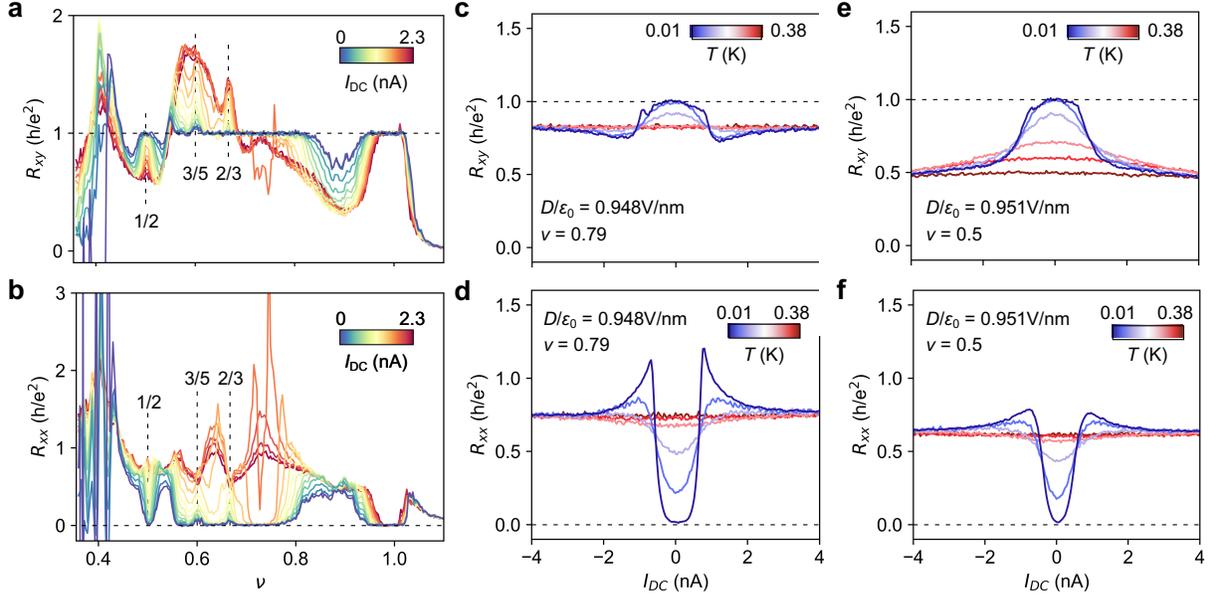

*Figure 3. Break-down of EQAH states in Device 1 the penta-layer graphene/hBN device. a&b. Differential resistance $R_{xx}$ and $R_{xy}$ taken at the dashed lines in Fig. 2a&b at 10 mK, using an AC current of 50 pA on top of a varied DC current. The quantized $R_{xy} = h/e^2$ and vanishing $R_{xx}$ can be seen at small DC current, while the high-temperature behavior (including FQAHE at v = 3/5 and 2/3) as shown in Fig. 2 is qualitatively reproduced at $I_{DC}$ = 2.3 nA. This suggests that the EQAH state is the true ground state at this displacement field in certain range of D, while the fractional quantum Hall liquid can be recovered by a large DC current. Non-monotonic dependence of $R_{xx}$ on $I_{DC}$, however, can be seen at intermediated $I_{DC}$. This is clearly different from the monotonic dependence of $R_{xx}$ on the temperature, as shown in Fig. 2c. c&d. Current-dependent $R_{xx}$ and $R_{xy}$ taken at (v, $D/\varepsilon_0$) = (0.74, 0.948 V/nm) and varying temperatures, revealing the quantized transport at small current in the EQAH state, and the anomalous quantum Hall liquid transport at high current when the EQAH state is broken down. The pair of peaks in $R_{xx}$ at the break-down threshold corresponds to the non-monotonic dependence of $R_{xx}$ on $I_{DC}$ in b, and is reminiscent of s-wave superconductors and Chern insulators to some extent. e&f. Current-dependent $R_{xx}$ and $R_{xy}$ taken at (v, $D/\varepsilon_0$) = (0.5, 0.951 V/nm) and varying temperatures, showing similar behaviors as in c&d.*

## Displacement Field-Induced Phase Transitions

The displacement field $D$ provides an important tuning knob of the flat band physics in rhombohedral graphene. With a small change of $D$, one can fine-tune the band structure to influence the competition between different ground states that are close in energy. We now examine the effect of $D$ on the observed EQAH states, as well as other electron liquid states in the same device.

Figure 4a&b show the $R_{xy}$ and $R_{xx}$ as a function of $D$. At around $D/\varepsilon_0 = 0.92$ V/nm, $R_{xy}$ shows a value of $2h/e^2$, which is compatible with the composite Fermi liquid (CFL) picture. At high temperatures, $R_{xy}$ decreases from $2h/e^2$ gradually while $R_{xx}$ non-monotonically as $D/\varepsilon_0$ changes from 0.9 to 0.94 V/nm. At low temperatures, a plateau at $R_{xy} = h/e^2$ and $R_{xx} = 0$ emerges at intermediate $D$s. This observation is consistent with the EQAH state in the diamond-shaped region at low temperatures, and indicates quantum phase transitions from the CFL to the EQAH state to the valley-polarized Fermi liquid (FL) triggered by $D$.

Figure 4c&d show $R_{xy}$ and $R_{xx}$ as a function of $D$ at varying DC current bias. Again, the plateaus at $R_{xy} = h/e^2$ and $R_{xx} = 0$ emerge at intermediate $D$ only when the $I_{DC}$ is small. Such dependence of the differential resistances on current does not show up for the CFL, as can be seen in Fig. 4e&f, where both $R_{xy}$ and $R_{xx}$ are constants regardless of the temperature. In contrast, in the range of $D$ corresponding to the EQAH state, the breakdown happens only at low temperatures but not high temperatures, as shown in Fig. 4g&h. The differences in response to temperature and current clearly mark the different nature of the likely-to-be CFL and the EQAH state.

At some fractional fillings where FQAHE were observed, $D$ can also induce quantum phase transitions. This is shown in in Fig. 4i-k, in which the filling factor is fixed at $v = 3/5, 4/7, 5/9$, respectively[48]. At the base temperature of 10 mK, FQAH state and EQAH state both occupy a range of $D$, as evidenced by the plateaus at $R_{xy} = h/ve^2$ and $R_{xy} = h/e^2$. Further tuning of $D$ can induce phase transitions to Fermi liquid and Wigner crystals at the high and low ends of $D$, respectively.

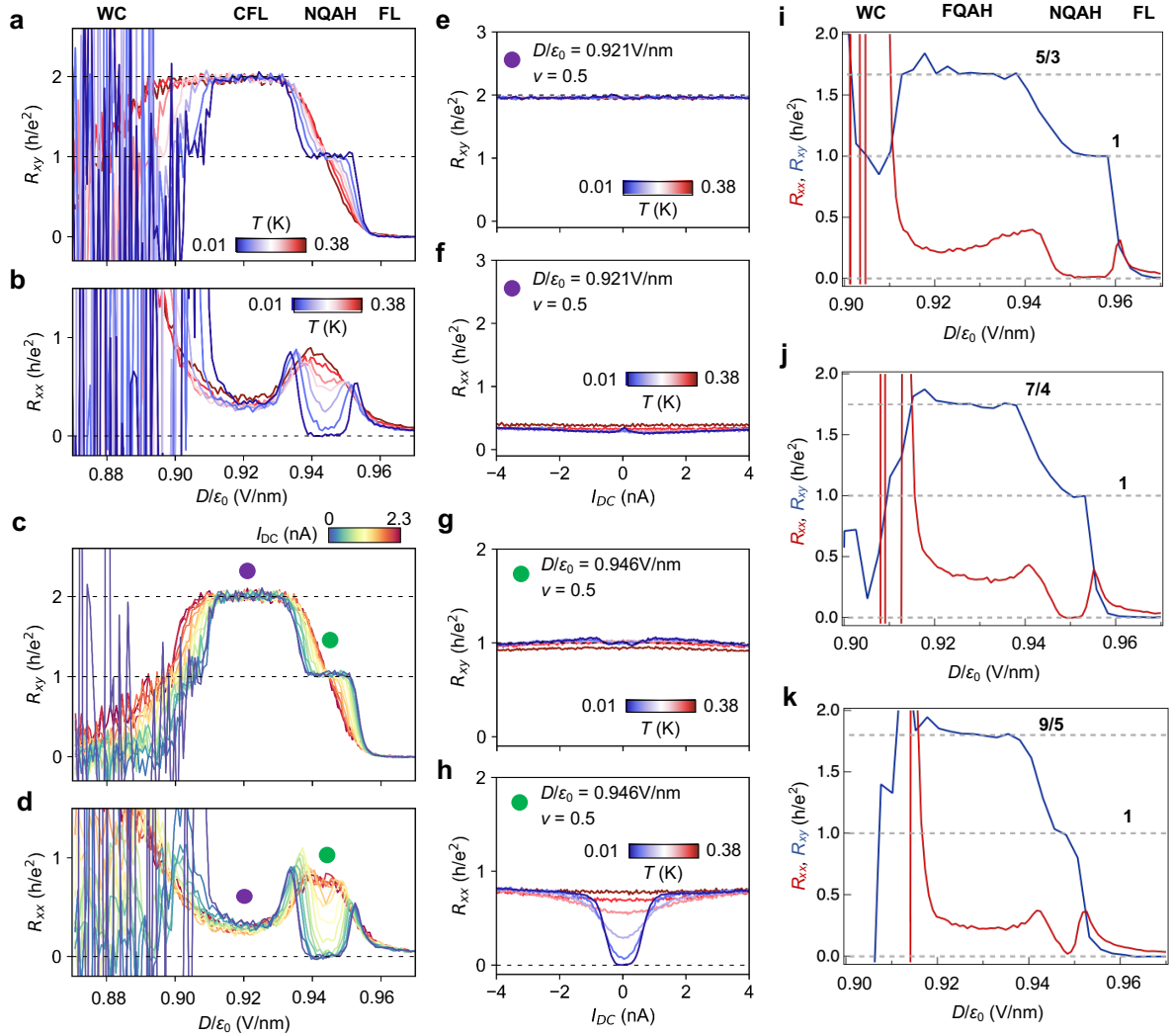

***Figure 4. Phase transitions from the EQAH states to (composite) Fermi liquid and FQAH liquid. a&b.*** $R_{xy}$ *and* $R_{xx}$ *at* $v = 1/2$ *and varying D in a range of temperatures. The plateaus at* $R_{xy} = 2h/e^2$ *and* $h/e^2$ *that correspond to CFL and EQAH can be clearly seen at low temperatures. At very high D, the transport behavior agrees with a valley-polarized Fermi-liquid. At very low D, the resistance diverges and suggests a Wigner crystal state.* ***c&d.*** *Differential resistances* $R_{xy}$ *and* $R_{xx}$ *taken at* $v = 1/2$ *and varying D at 10 mK, using an AC current of 50 pA on top of a DC current. Behaviors in different ranges of D qualitatively agree with that in* ***a&b***. ***e&f.*** *Current dependence of differential resistances* $R_{xy}$ *and* $R_{xx}$ *at* $(v, D/\varepsilon_0) = (0.5, 0.921 \text{ V/nm})$, *agreeing with the linear voltage-current transport of CFL.* ***g&h.*** *Current dependence of differential resistances* $R_{xy}$ *and* $R_{xx}$ *at* $(v, D/\varepsilon_0) = (0.5, 0.946 \text{ V/nm})$, *revealing the linear voltage-current transport behavior at high temperatures when the EQAH state is broken-down to fractional quantum Hall liquid, which is contrasted by the non-linear voltage-current transport when the EQAH is un-perturbed at low temperatures.* ***i-k.*** *Dependence of* $R_{xy}$ *and* $R_{xx}$ *on D at 10 mK, taken at* $v = 3/5, 4/7$ *and* $5/9$ *respectively. Plateaus corresponding to FQAH and EQAH states are clearly seen and connected by phase transitions induced by D at the same v.*

## Discussion

Our observation of quantized $R_{xy}$ and vanishing $R_{xx}$ in a wide range of $v = 0.5$ to up to 1.3 at zero magnetic field, the temperature dependences of resistances, and the strong threshold behavior in transport at small current indicate that the EQAH state is a new topological phase of correlated electrons. Here we discuss more on the nature of this EQAH state with two possible underlying pictures.

In picture 1, what is observed could be similar to the quantum anomalous Hall crystal (QAHC) suggested by several recent theory works on the rhombohedral graphene[3,4,11,12]. QAHC could exist in the absence of moiré effect and would break the time-reversal-symmetry and translational symmetry spontaneously and simultaneously. The experimentally observed EQAH prevails in a wide range of electron density and moiré filling factor, and most of them are in-commensurate with the moiré superlattice. This is distinct from the generalized Wigner crystal which only shows up at fractional filling factors of the moiré superlattice[49–52]. Therefore, the role of moiré in the formation of the EQAH state is likely different from that in the formation of generalized Wigner crystals (or charge density wave). It is possible that the weak but non-zero moiré potential (through a remote Coulomb potential, for example) could facilitate the formation of QAHC by modulating the band structure and Berry curvature distribution. Experimentally, an ideal experiment to test the QAHC picture is to measure rhombohedral graphene devices in the absence of moiré effect (for example, at large twist angles between graphene and hBN). We note that in order to stabilize the QAHC in the absence of moiré, it is likely that even lower electronic temperatures, smaller bias current, and the right amount of impurities (to pin the electron crystal as in the case of Wigner crystals) will be needed. These conditions are to be carefully engineered in future experiments.

In picture 2, the observed EQAH state could be similar to the re-entrant quantum Hall effect (RQHE) state in two-dimensional electron gas at high magnetic fields[53–59]. The RQHE can be generally understood as the superposition of a gapped integer quantum Hall liquid[17] at integer filling of Landau levels and a Wigner crystal formed by the excess charges. In analogy, topological flat bands play the role of Landau levels and the integer QAHE can be contributed by a QAH liquid at $v = 1$. The charges corresponding to $(1-v)$ will have to form a Wigner crystal that contributes a diverging $R_{xx}$ and zero $R_{xy}$. In this case, although it is not entirely clear how moiré superlattice drive the QAHE at $v = 1$, some kind of moiré effect is needed for the formation of EQAH states as we observed. However, we note that the RQHE happens predominantly in high-index Landau levels (with rare exceptions [54,59]), while the EQAH state we observed happens mostly in the lowest moiré band. RQHE was also never observed for a large and continuous filling range as we observed for the EQAH states. Especially, RQHE never kills fractional quantum Hall states, while EQAH states replace some FQAH states at the base temperature.

In either case, our observed EQAH state is a new topological phase of electrons that is distinct from other topological states so far: 1. It happens spontaneously at zero magnetic field and is thus different from all states exhibiting quantum Hall effects under a magnetic field; 2. It happens in a wide range of moiré filling factor and is thus different from all Chern insulators that are tied to specific integer or fraction fillings. Both the two pictures above and the threshold behavior of electric transport suggest an underlying electron crystal in the EQAH state. In two-dimensional electron gases, topologically trivial electron crystals have been realized when the effective mass is large[60–63], by quenching the kinetic energy under high magnetic fields[64–69], or be stabilized by the commensuration with an underlying moiré superlattice or another layer hosting electrons[49–51]. A topological electron crystal at zero magnetic field, however, has never been observed in any experiment. Further experiments are needed to test the possible electron crystal nature of our observed EQAH state, through direct imaging in the real space[50,62,69], measuring electron diffraction peaks in the momentum space[70], or by measuring narrow-band noise[71–73]. These measurements would require more advanced device engineering, better control of electronic temperature, and novel microscopy experiments, which are beyond the scope of this work. In the meanwhile, our experiment opens up opportunities to study the QAHC, an unprecedented topological state of matter that breaks both the time-reversal and translational symmetries spontaneously.

# Method

**High Through-Put Fabrication of Rhombohedral Graphene Stacks Enabled by Advanced Infrared Imaging and Device Fabrication**

Device 1&2 were used in Ref.[1] where the description of device fabrication can be found. Device 3 was made in generally the same procedures as Device 1&2, except for the imaging of the rhombohedral stacking order. We developed a new infrared imaging technique based on InGaAs camera, which is installed on a regular optical microscope and can take pictures of the graphene flakes. Different stacking orders in graphene can be identified based on their different infrared conductivities and contrast with the substrate, as described in Ref[74,75]. Compared to the near field infrared nanoscopy technique we employed previously, the far-field imaging based on the InGaAs camera has a much higher efficiency due to the multi-pixel data collection simultaneously. We combine the near-field and far-field approaches, together with Raman spectroscopy to visualize and confirm stacking orders in the exfoliated flakes and assembled stacks.

Extended Data Figure 1 shows how this imaging method is implemented. We picked up the top hBN, graphite, middle hBN, and the penta-layer graphene using polypropylene carbonate (PPC) film and landed it on a prepared bottom stack consisting of an hBN and graphite bottom gate. We used the IR camera to quickly screen the exfoliated graphene flakes and identify the rhombohedral domains. We also checked if the stacking order is preserved after picking up the graphene with hBN and after the stack is finished. The device was then etched into a Hall bar structure using standard e-beam lithography (EBL) and reactive-ion etching (RIE). We deposited Cr/Au for electrical connections to the source, drain and gate electrodes.

There are two caveats to note though, as shown in Extended Data Figure 2. Firstly, for multilayer graphene thicker than three layers, there are more intermediate stacking orders other than rhombohedral and Bernal stacking. Some of the intermediate stackings show similar responses to the rhombohedral stacking on the InGaAs camera and could mislead the judgment. Near-field infrared nanoscopy works better in differentiating rhombohedral stacking from other stackings, possibly due to the longer wavelength used in the near-field measurements. Secondly, the spatial resolution of the far-field imagining is limited to ~one micron due to the diffraction limit of infrared light, and could miss domain structures with mixed stacking orders and domain walls that are smaller than the spatial resolution[32,76].

**Transport measurement**

The device was measured in a Bluefors LD250 dilution refrigerator with an electronic temperature of around 40 mK. Electronic temperature is estimated by a superconducting state in other graphene device with Tc around 40mK as shown in extended Figure 3. Stanford Research Systems SR830 lock-in amplifiers were used to measure the longitudinal and Hall resistance $R_{xx}$ and $R_{xy}$ with an AC frequency at 17.77 Hz. The DC and AC currents are generated by Keysight 33210A function generator through a 300MOhm resistor. Keithley 2400 source-meters were used to apply top and bottom gate voltages. Top-gate voltage $V_t$ and bottom-gate voltage $V_b$ are swept to adjust doping density $n_e = (C_t V_t + C_b V_b)/e$ and displacement field $D/\varepsilon_0 = (C_t V_t - C_b V_b)/2$, where $C_t$ and $C_b$ are top-gate and bottom-gate capacitance per area calculated from the Landau fan diagram.

**Dis-entangling longitudinal and Hall resistance**

Devices 1&2 are the same as used in Ref.[1] and has been measured in the same way to dis-entangle the $R_{xx}$ and $R_{xy}$. As shown in Fig. 1c, device 3 is in a Hall bar geometry with good electric contacts on both sides of the channel at high displacement fields. We use electrodes 'S' & 'D' to pass the current, electrodes '1' & '2' to measure $R_{xx}$ and electrodes '2' & '3' to measure $R_{xy}$. Measurements performed at opposite magnetic fields (larger than the coercive field) can thus be used to extract $R_{xx}$ and $R_{xy}$ at $B = 0$ for the IQAH, FQAH and EQAH states, following:

$R_{xx}(0) = (R(B) + R(-B))/2$      and      $R_{xy}(0) = (R(B) - R(-B))/2$.

**EQAH States in Device 2**

Extended Data Figure 5 shows data from Device 2, the other penta-layer graphene/hBN device on the same chip as Device 1. The twist angle and general phase diagram of Device 2 are similar to those of Device 1, featuring FQAHE, EQAH states and phase transitions similar to described in Fig. 1-4 in the main text.

**FQAHE and EQAH States in Device 3**

Extended Data Figure 6-8 show data from Device 3, a newly made tetra-layer rhombohedral graphene/hBN device. At 300 mK, the phase diagram shows IQAHE and FQAHE. At 10 mK, the EQAH states emerge and dominate a large range of ($v$, $D$) from $v = 0.5$ to 1.3.

**Author Contributions**

L.J. supervised the project. Z.L. and T.H. performed the DC magneto-transport measurement. T.H., Y.Y., L.S., and S.Y. fabricated the devices. J.Y., J.S., Z.L. and T.H. helped with installing and testing the dilution refrigerator. K.W. and T.T. grew hBN crystals. All authors discussed the results and wrote the paper.

**Competing Interests** The authors declare no competing interests.

**Data availability** The data that support the findings of this study are available from the corresponding authors upon reasonable request.

**Extended Data Figures**

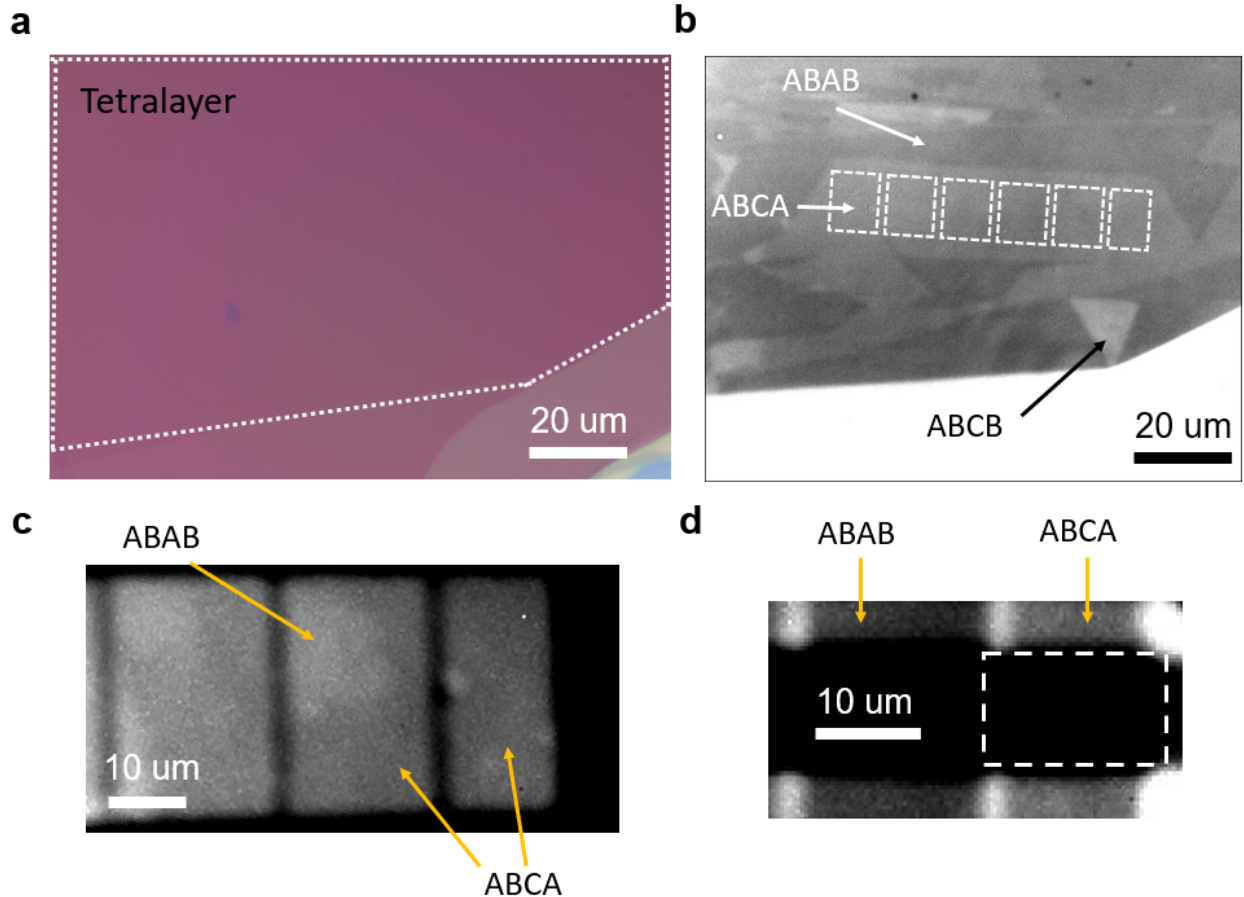

***Extended Data Figure 1. Visible and infrared far-field imaging of multilayer graphene during fabrication. a.*** *Optical image of an exfoliated graphene flake on SiO$_2$/Si substrate, taken with a regular optical microscope. The majority of the flake (outlined by dashed lines) is tetra-layer. Scale bar: 20 μm. No contrast can be seen within the tetra-layer region.* ***b.*** *Infrared image of the same flake as in **a**, taken by an InGaAs camera. Domains with different contrasts can be seen, which correspond to different stacking orders. The rhombohedrally stacked domain (ABCA) has a contrast between the Bernal stacked domain (ABAB) and an intermediate stacked domain (ABCB). These stacking orders have been confirmed by Raman scattering measurement. We cut 6 rectangles out of the ABCA domain using a laser, as outlined by dashed boxes.* ***c.*** *The InGaAs camera image of 3 (out of 6 ABCA) flakes that are picked up. The two flakes on the left have partially changed to other stacking orders, while the right-most flake remains in ABCA stacking.* ***d.*** *The InGaAs camera image of the three flakes in **c** after they are dropped down to hBN and bottom graphite gate. The right-most flake in **c** remained in the ABCA stacking, as indicated by*

*the right arrow. It has a clear contrast with the middle flake in **c**. The ABCA flake in the dashed box was made into Device 3.*

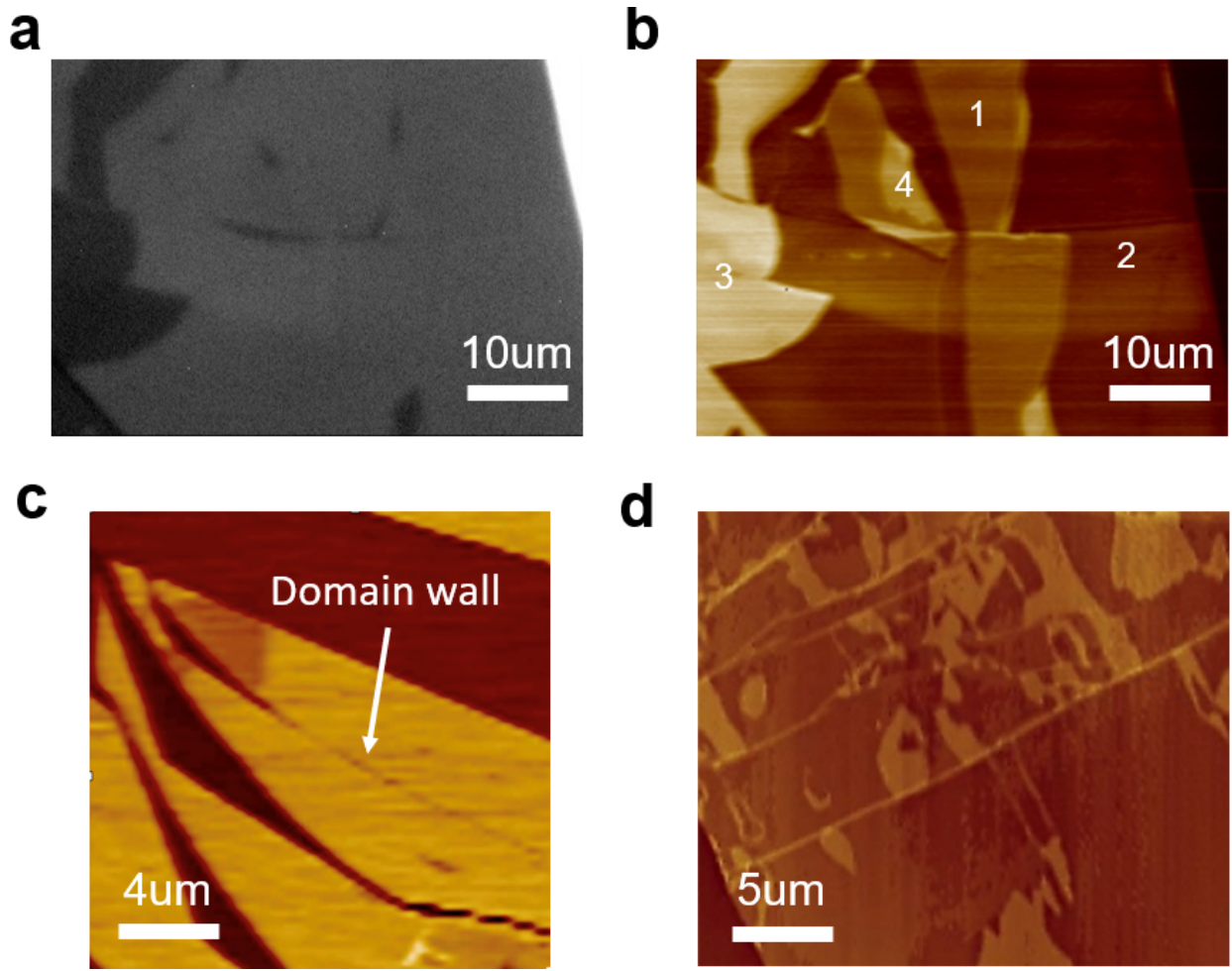

***Extended Data Figure 2. Comparison between far-field infrared imaging and near-field infrared imaging. a&b.** InGaAs camera image and near-field infrared nanoscopy image of an exfoliated multilayer graphene flake on SiO$_2$/Si substrate. The latter reveals more domains (labeled as 1-4) with clear contrast than the former does. **c&d.** Near-field infrared nanoscopy images of two exfoliated trilayer graphene flakes. Both images show domain and domain walls that have dimensions much smaller than 1 μm, which is well-below the diffraction-limit of the far-field imaging based on the InGaAs camera.*

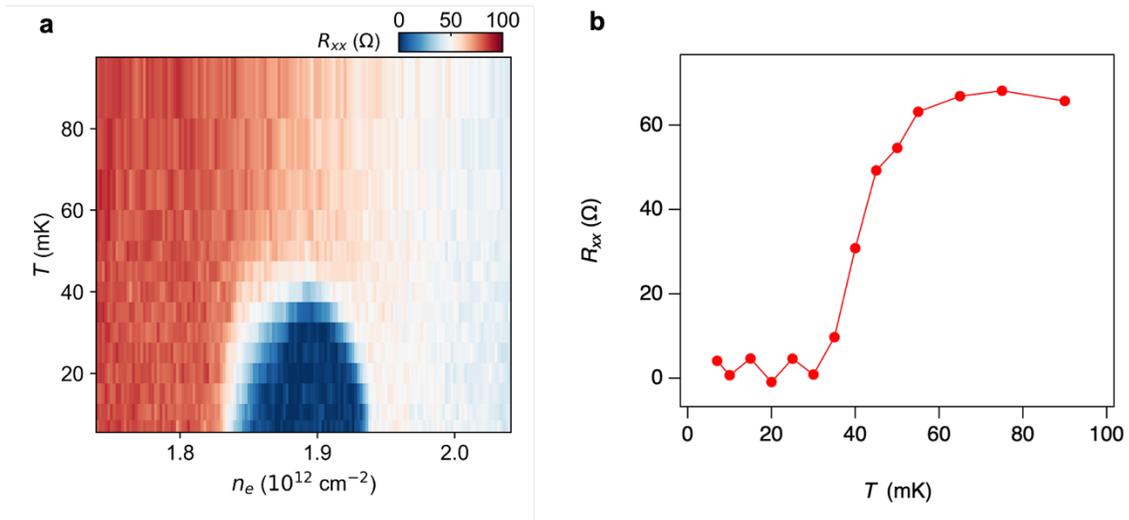

***Extended Data Figure 3. Estimation of the electron temperature. a.*** *Temperature dependence of $R_{xx}$ in a graphene superconducting state. The density range of the superconducting dome remains expanding below 40mK.* ***b.*** *Temperature dependence of $R_{xx}$ at $n_e = 1.9 \times 10^{12}$ cm$^{-2}$.*

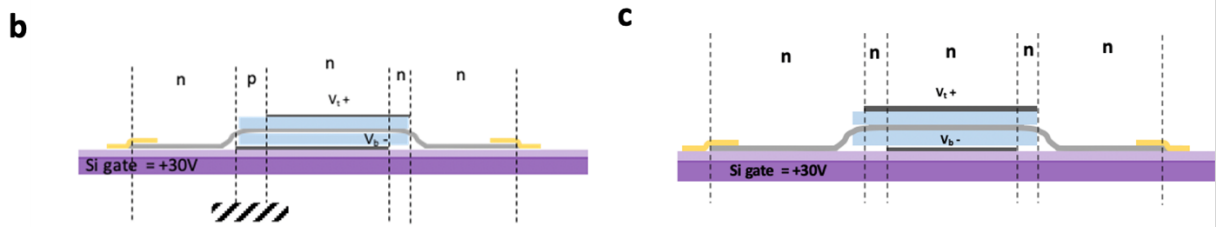

*Extended Data Figure 4. Summary of the device and contacts information. **a.** Basic information of the three devices. **b.** Schematics of the gates and contacts layout in device 1 and 2. Top and bottom graphite gate shifted relative to each other creating a n-p-n junction on one side of the contacts. **c.** Gates and contacts layout in device 3 with optimized geometry. No p-n junction will be formed on either side of the device.*

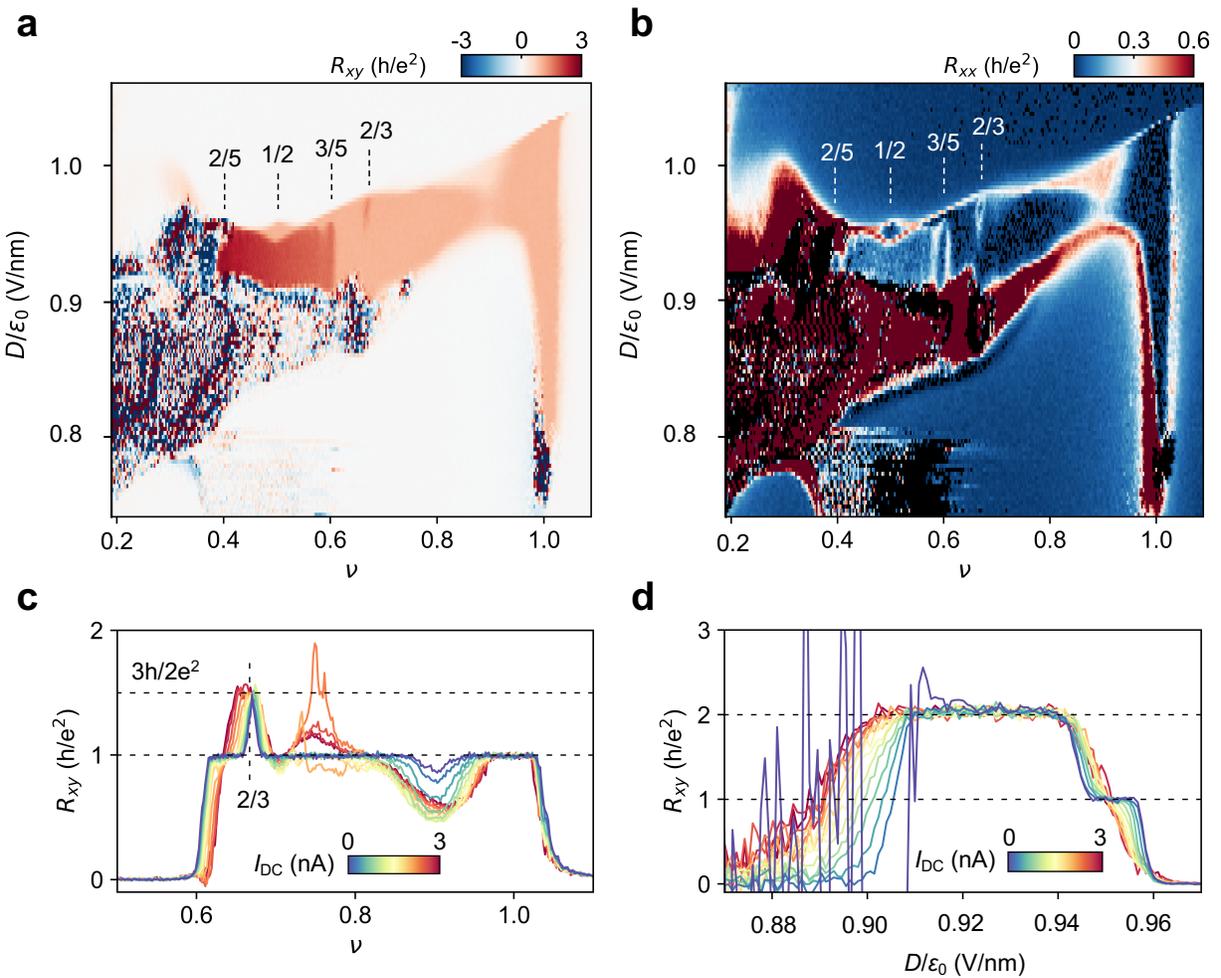

*Extended Data Figure 5. Phase diagram of Device 2, a penta-layer rhombohedral graphene/hBN moiré superlattice, at $I_{DC}$ = 0 A and $I_{AC}$ = 0.2 nA. **a&b.** Mapping of $R_{xx}$ and $R_{xy}$ in a large range of $v$ and $D$ at a mixing chamber temperature of 10 mK. The main features are similar to that of Device 1 shown in Fig. 2. Three regions show quantized $R_{xy}$ at $h/e^2$ and vanishing $R_{xx}$, located at around $v = 1/2$, spanning between $v = 0.55$ and $0.83$, as well as around $v = 1$. These*

three regions are almost connected into one big region with $R_{xy} = h/e^2$ that swamps the FQAH states at $v > 1/2$. **c.** Line-cut of $R_{xy}$ at $D/\varepsilon_0 = 0.96$ V/nm, featuring a wide plateau at $h/e^2$. **d.** $R_{xy}$ at $v = \frac{1}{2}$, featuring the phase transition from CFL to EQAH driven by displacement field.

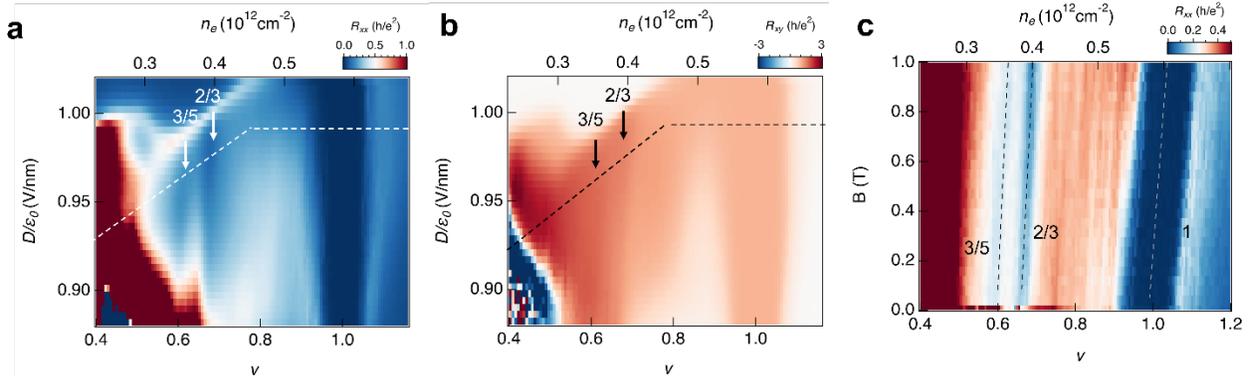

***Extended Data Figure 6. FQAHE in Device 3 at 300 mK. a&b.*** *Mapping of $R_{xx}$ and $R_{xy}$ in a large range of v and D at $I_{DC} = 0$ nA and $I_{AC} = 0.1$ nA, which is beyond the break-down threshold current of EQAH states. **c.** Landau fan corresponding to the line-cut in **c**, where the dashed lines are derived from the Streda's formula for states with $R_{xy} = 5h/3e^2$, $3h/2e^2$ and $h/e^2$, respectively. The dispersions of dips in $R_{xx}$ agree well with the dashed lines, as expected for fractional and integer Chern insulators at $v = 3/5$, $2/3$ and $1$.*

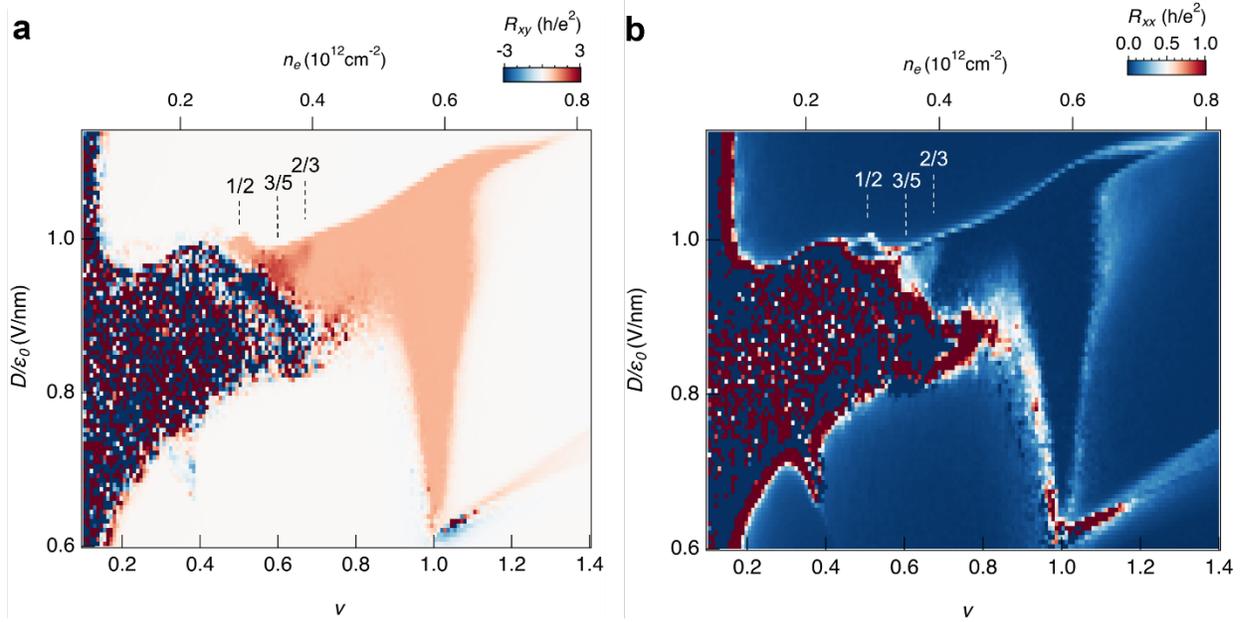

*Extended Data Figure 7. Phase diagram of Device 3, a tetra-layer rhombohedral graphene/hBN moiré superlattice, at 10 mK, $I_{DC}$ = 0 A and $I_{AC}$ = 0.2 nA. a&b. Mapping of $R_{xx}$ and $R_{xy}$ in a large range of v and D at a mixing chamber temperature of 10 mK. Different from Device 1&2, the region with quantized $R_{xy}$ at $h/e^2$ and vanishing $R_{xx}$ shifts to higher moiré filling factors. This is likely due to the smaller twist angle and smaller charge density corresponding to v = 1 in Device 3. As a result, region 2 and 3 in Fig. 2 merge into one region without a gap in between. At the same time, the EQAH region extends significantly beyond v = 1 and reaches v = 1.3.*

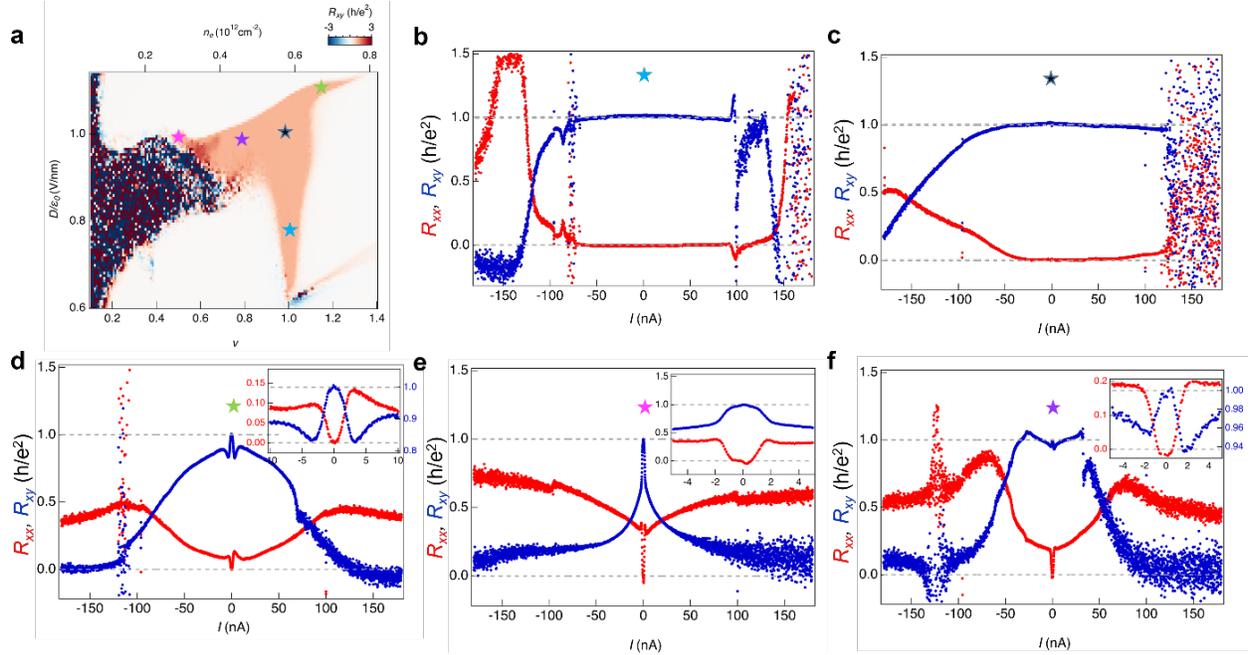

***Extended Data Figure 8. Current-induced break-down of EQAH state and Chern insulators in Device 3 at 10 mK. a.*** *Mapping of $R_{xy}$ in a large range of v and D at $I_{DC}$ = 0 A and $I_{AC}$ = 0.2 nA. Five states at 'stars' positions show quantized $R_{xy}$ at $h/e^2$ and vanishing $R_{xx}$.* ***b&c.*** *$R_{xx}$ and $R_{xy}$ as a function of $I_{DC}$ at the light and dark blue 'star' positions corresponding to v=1 in **a**, showing the break-down behavior of the C=1 Chern insulator at large current (>50 nA).* ***d-f.*** *$R_{xx}$ and $R_{xy}$ as a function of $I_{DC}$ at the green, pink and purple 'star' positions in **a**, showing the break-down behavior of the EQAH states at small current (< 1 nA) in contrast to the large break-down current of the C=1 Chern insulator in **b&c**. Inset: zoom-in of curves at around $I_{DC}$ = 0 nA, which reveal the threshold of EQAH break-down at ~ 1 nA. Note: Spike-like rapid changes in figure b-f are artifacts from voltage source meter when switching output range.*